\documentstyle[12pt]{article}
\begin{document}
\def\etal{{\em et al.\ }}
\def\etaten{\eta_{10}}
\def\he{${{}^4He\:\:}$}
\def\be{\begin{equation}}
\def\ee{\end{equation}}
\def\bc{\begin{center}}
\def\ec{\end{center}}
\def\beq{\begin{eqnarray}}
\def\eeq{\end{eqnarray}}
\def\beqd{\begin{eqnarray*}}
\def\eeqd{\end{eqnarray*}}
\def\nin{\noindent}
\def\non{\nonumber}
\def\lra{$\leftrightarrow$ }
\def\eset{$\not\!\!\:0$ }
\def\bull{$\bullet$}
\def\svac{\sin^2 2\theta_V}
\def\MC{Monte-Carlo }
\def\ML{Maximum Likelihood }
\def\LL{Lamb and Loredo }
\def\Kam{Kamiokande }
\def\SN{SN1987A }
\def\mnueb{m_{\bar{\nu}}_e}
\def\mnue{m_{\nu_e}}
\def\tnueb{T_{{\bar{\nu}}_e}}
\def\tnuebz{T_{{\bar{\nu}}_e}^0}
\def\tnue{T_{\nu_e}}
\def\tnuez{T_{\nu_e}^0}
\def\tnuezm{{\cal T}_{\nu_e}^0}
\def\tnum{T_{\nu_{\mu}}}
\def\tnumb{T_{\bar{\nu}_{\mu}}}
\def\tnut{T_{\nu_{\tau}}}
\def\nueb{{\bar{\nu}}_e}
\def\nue{\nu_e}
\def\num{\nu_{\mu}}
\def\numb{{\bar{\nu}}_{\mu}}
\def\mnue{m_{\nu_e}}
\def\tkii{t_{off}^{kam}}
\def\timb{t_{off}^{imb}}
\def\unlikem{(\mnue)_{L^-}}
\def\likem{(\mnue)_{L^+}}
\def\unlikekii{(\tkii)_{L^-}}
\def\likekii{(\tkii)_{L^+}}
\def\unlikeimb{(\timb)_{L^-}}
\def\unlikemnue{(\mnue)_{L^-}}
\def\likemnue{(\mnue)_{L^+}}
\def\likeimb{(\timb)_{L^+}}
\def\liketau{(\tau_c)_{L^+}}
\def\likealpha{(\alpha)_{L^+}}
\def\liketnu{(\tnuez)_{L^+}}
\def\liketnub{(\tnuebz)_{L^+}}
\def\ptoffdet{P(t_{off}^{det} < t |N^{det},\mnue)}
\def\ptoffkam{P(t_{off}^{kam} < t |N^{kam},\mnue)}
\def\ptoffimb{P(t_{off}^{imb} < t |N^{imb},\mnue)}
\def\lnL{\ln {\cal L}}
\def\Ebind{E_{53}^B}
\def\omb{\Omega_B}
\pagestyle{empty}
\rightline{{\bf CWRU-P7-94}}
\rightline{SEPT 1994}
\baselineskip=16pt
\begin{center}
\bf\large {YET ANOTHER PAPER ON SN1987A: LARGE ANGLE OSCILLATIONS,
AND THE ELECTRON NEUTRINO MASS}
\end{center}
\vskip0.2in
\begin{center}
Peter J. Kernan and Lawrence M. Krauss\footnote{Also Dept of Astronomy}
\vskip .1in
 {\small\it Department of Physics\\
Case Western Reserve University\\ 10900 Euclid Ave., Cleveland, OH
44106-7079}
\vskip 0.4in
\end{center}

\begin{center}
{\bf Abstract}
\end{center}

\baselineskip=16 pt

We supplement \ML methods with a \MC simulation to
re-investigate the SN1987A neutrino burst detection by the IMB and
Kamiokande experiments. The detector simulations include background in the
the latter and ``dead-time'' in the former.  We
consider simple neutrinosphere cooling models, explored previously in
the literature, to explore the case for or against neutrino vacuum mixing
and massive neutrinos. In the former case,
involving kinematically irrelevant masses, we find that the full range of
vacuum mixing angles, $0 \le \svac \le 1$, is permitted, and the \ML
mixing angle is $\svac = .45.$ In the latter case we
find that the inclusion of ``dead-time'' reduces previous $\mnue$
upper bounds by 10\%, and supplementing the \ML analysis with a
\MC goodness-of-fit test results in a further 15\%
reduction in the $\mnue$ upper limit. Our 95\% C.L. upper limit for
$\mnue$ is 19.6eV, while the best fit value is
$\sim$ 0eV.

\newpage
\pagestyle{plain}
\baselineskip=21pt

\nin {\bf Introduction }

Galactic neutrino astronomy began in 1987 with the
observation of 20 neutrinos from the supernovae burst \SN in the
Large Magellanic Cloud (LMC). Two terrestial detectors, IMB \cite{imb}
and Kamiokande \cite{kam},
found unequivocal evidence for supernovae neutrino events with the former
collaboration claiming detection of 8 \SN events, and the latter 12. This
momentous observation generated enormous excitement in the scientific
community, and of course a plethora of papers soon followed. We have
returned to this subject for three reasons. First,
apparently only the IMB collaboration took into account
the ``dead-time'' in the IMB detector when comparing these
observations with theory. Also we believe that a more or less
model-independent approach to the question of vacuum mixing in \SN
should be done (and with more rigor than in \cite{krauss:nature}).
Finally, we have in hand a detailed Monte Carlo code for generating
the predicted signal in light water detectors which was created for the
purpose of exploring the nature of a galactic neutrino signal
\cite{krauss:nucb}, but which can also be used to accurately model the
signal for SN1987A.  Using this code we felt that it might be possible
to improve upon the neutrino mass limits derived previously for the
SN1987A neutrino burst.

Before describing our analysis, it is worthwhile briefly reviewing the
most recent results on neutrino mass and mixing constraints from
SN1987A in order to comment on the improvements incorporated in our
present analysis.  To date, the most comprehensive statistical analysis
of the signal was performed by Lamb and Loredo
\cite{ll}, who used a Maximum Likelihood technique,
and were the first to incorporate the background event rate in a
likelihood function in order to account for at least one of the
Kamiokande events which was probably a background event (which
several previous analyses had simply discarded).  To facilitate
comparison with their results (in the case of a significant neutrino
mass) we exploited essentially the same formalism, but made several
minor additions. Primarily, we
included the fact that the IMB detector had a significant amount of
``dead-time'' ($\sim 13\%$ over theburst
\cite{imb}) following interactions in the detector by cosmic ray muons
or \SN neutrinos. The reason for the large dead-time,
35ms/interaction, is due to the data-acquisition software \cite{sd}.
This 35ms dead-time would of course not be problematic for the
original purpose of the IMB detector, measuring proton decay.

A few investigators \cite{ll,walker} allowed
 {\it independent}
offset times (the time-delay between the first \SN neutrino arriving at the
earth and
the first detection in either \Kam or IMB) for each detector.
We noticed  that
in the case of a significant neutrino mass,
when the offset times, $t_{off}$,
 are important,
the \ML value $\tkii$ (\Kam offset time) at the $\mnue$ upper limit of
Lamb and Loredo seemed unacceptably long. This suggested extending the \ML
approach.  The
\ML method does not test a model, but rather tests the allowed range of
parameters given a specific model. Thus should the
\ML method locate a parameter value which may seem otherwise
unlikely one must use other methods, such as a Monte-Carlo, to test
the model.  Here we were able to exploit the power of the
comprehensive light water neutrino detector Monte Carlo code
previously written to examine various features of possible future
galactic neutrino burst signals
\cite{krauss:nucb}. This code incorporates all aspects of the detector in
order to generate a realistic signal, given a specific supernova burst
model.  Using this Monte Carlo, we can show, as we describe in more
detail below, that the
\LL
\ML offset time for
\Kam of 3.9 sec would be expected to occur in less than 1 out of 400
cases given the other \ML
\SN parameters, such as binding energy, emission timescale, etc.

We next turn to the issue of neutrino mixing constraints. There has
recently been a model-dependent derivation of bounds on vacuum
oscillations from the
\SN data \cite{smirnov}, with 5 specific explosion models
considered.  All models fit the no-mixing scenario and, perhaps not
surprisingly, when the amount of neutrino mixing was increased a
Kolmogorov-Smirnov test led to a lower bound of $\svac < .7 \to
.9$. The upper bound of .7 would exclude the ``just-so" solution
\cite{justso} to the solar neutrino problem \cite{snp} and
much of the large angle region of the MSW solution. The severity
of this bound is surprising, given the sparsity of the observed
signal, so we decided to examine this issue in some more detail.  The
authors of
\cite{smirnov} recognized the fact that their result was model
dependent, but just how model dependent was not clear.  The ability to
explore the neutrino signal with our Monte Carlo makes it very easy to
sample supernova model space. We will show below that with a minimally
model-dependent approach, maximal vacuum mixing actually fits the data
better (greater likelihood) than no mixing.

Finally, we note that the neutrino mass limit we derive here has
already been superceded by direct laboratory probes \cite{pdg}.
Nevertheless, the utility of exploiting a galactic supernova burst to
constrain neutrino masses and mixings remains of great interest, and
the techniques we examine here thus remain important to explore.
Namely, SN1987A remains, even 7 years later, an important, and unique
test case if we are to attempt to fully exploit the information which
may be available in the next observed supernova neutrino burst.

\vskip .1in
\nin {\bf The Minimal Model }
\vskip .1in

We exploit here two ``minimal" models. The version we use for
bounding the neutrino mass and the one used for bounding the vacuum
mixing angle differ in that the latter has an extra time constant in the
neutrino spectra, which we will discuss later. In the former case we
follow \LL and assume a simple exponential cooling model. In this case
the supernovae is characterized by a Fermi-Dirac neutrino spectrum and 3
parameters; a maximum initial temperature $\tnuebz$, a cooling
time-scale $\tau_c$, and $\alpha$, related to the size of the
neutrinosphere,

\be
\alpha = {R_{10} \over D_{50}}.
\ee

\nin With $R_{10}$ the radius of the neutrinosphere in units of 10
kilometers and $D_{50}$ the distance to the LMC in units of 50
kiloparsecs. Alternatively one can view $\alpha$ as a relation for the
supernovae binding energy with the additional assumption that there is
an equipartion of the binding energy carried away among the 3 flavor
states times 2 spin states of the emitted neutrinos.

\be
E_{53} = 3.39 \times 10^{-4} \alpha^2 D_{50}^2 \int {\tnueb}^4(t) dt
\ee

\nin with $E_{53}$ the neutron star binding energy in
units of $10^{53} ergs$ and $\tnueb$ given in the simple exponential
cooling model by

\be
\tnueb(t) = \tnuebz \exp(-t/4 \tau_c).
\label{eq:tnu}
\ee

The model is also characterized by the neutrino mass, $\mnue$, and an
additional parameter for each detector, $t_{off}$, the offset
time. The offset time is particularly important for the massive
neutrino case where the neutrino mass causes a delay-time in the
arrival of the neutrinos:

\be
\Delta t = 2.57 (\mnue^2)_{eV^2} E_{MeV}^{-1} D_{50} s,
\ee

\nin with $\mnue$ in units of eV and $E_{MeV}$,
the incident neutrino energy, in MeV. Independent offset times are needed
for each detector \cite{walker} because of a problem with the \Kam clock
during the time that the \SN neutrino burst passed the earth. The
offset times play a major role in constraining
non-zero neutrino masses, due to the difficulty, when maximizing the
likelihood function for the \Kam detector, of reconciling a few early
low energy neutrino events, which would then imply a large offset time
for non-zero mass, with following high energy events, which
would favor a small offset time
\cite{kolb}.

The second version of our minimal model, in the case of vacuum mixing
and nearly massless neutrinos, introduces an additional parameter for the
anti-neutrino temperature,
$\tnueb$.  As far as kinematics are concerned, in this part of
the analysis we assume effectively massless neutrinos
 (the time delays introduced by the very small masses of interest
in this case are irrelevant), and we assume two state mixing,
$\nue$ and
$\num$. Vacuum mixing implies the neutrino spectrum at the earth is a
mixture of the 2 original spectra according to

\be
d^2N^{inc}/dE dt = (1- .5\svac) d^2N^{\nueb} /dE dt
+ .5\svac d^2N^{\numb} /dE dt
\ee

Qualitative arguments suggest that $\tnum \simeq 2\tnue$ due to the fact
that the $\nue$ have additional interactions from neutrons and charged
current interactions, while $\num$ has only neutral-current interactions in
the supernovae environment.  Thus the $\num$ are emitted from deeper in the
star and hence their spectrum is characterized by a hotter temperature.  The
temperature evolution of the $\nueb$ is potentially slightly more
complicated: initially before neutronization of the star $\tnueb \simeq
\tnue$ is expected.  However as the star neutronizes and the reaction
$\nueb + p
\to n + e^+$ becomes rare then $\tnueb \to \tnum$ is expected
\cite{tnubar}.  Our model parameterizes this phenomenon by introducing
an additional time-constant which smoothly and symmetrically takes $\tnueb$
from $\tnue$ to $\tnum = 2\tnue$ in time $2\tau_2$.

\be
\tnueb = \left\{1. +
.5 \tanh\left[ (t - \tau_2 ) \pi /\tau_2)\right] \right\} \tnue(t)
\ee

\nin This function is constructed so that
when {\em t} advances to $\tau_2$ we have $\tnueb = .5(\tnue +
\tnum)$.
Introducing the extra parameter, $\tau_2$, allows us to
take a conservative approach to the consideration
of vacuum oscillations.
Also we take $\tnue$ and $\tnum$ to have the form
of eq.\ref{eq:tnu} (such that
if the \ML $\tau_2 \gg \tau_c$ this will reduce
$\tnueb$ to the
form used for massive neutrinos).

Another point worth mentioning here is that we continue to
partition the binding energy equally among all species here so that
the factor $\alpha$, which sets the scale of the fluxes,

\[ d^2N/ dE dt \propto \alpha^2 \]

\nin is different for the 3 species,
$\nue, \nueb $ and $\num$. These are related by,

\be
\alpha_{\nu_i}  =  \alpha_{\nue} \sqrt( \int \tnue^4 dt /
\int T_{\nu_i}^4 dt)
\ee

\nin with $\nu_i=\nueb,\num$.

As we have indicated, an improvement in our \ML method compared to the
\LL work is the inclusion of dead-time for the IMB detector.  The reason
one expects this may have an effect is due to the fact that the
\Kam data favors a cooler, less energetic supernovae than does the IMB
data. Our \MC work indicates that if one uses the IMB data to locate a
set of \SN parameters, these same parameters typically predict many more
events in \Kam $\sim 20$ , with a much higher average energy $\sim 22
MeV$, than were seen ( 12 events and 15 MeV, respectively).  Thus one
of the reasons why the \ML analysis can provide a reasonably localized
parameter space for the combination of the IMB and \Kam data is the
tension between the fits for the two separate data sets. Since including
the dead-time in IMB will favor an even more energetic supernovae it
should exacerbate the existing tension leading to stronger constraints on
parameters derived from the
\ML method.  One of the purposes of this letter is to show how
much the 13\% dead-time changes our conclusions from those of \LL.

The dead-time, $t_d=35 ms$, was handled in different ways in the \MC and \ML
methods. In the prior case it is straightforward. We use a Poisson
distributed random number generator to simulate the known 2.7Hz muon
event rate (this requires a 3Hz incident rate since dead-time affects
this measurement as well) starting at $t-1 sec$.
With the \MC neutrino events and muon
events in temporal order we then remove any neutrino events occurring
within 35ms of a previously ``detected" muon or neutrino.

For the \ML method, we modify the spectral rate according to,

\be
d^2N/dE dt \to (1 - P_d(t-t_d,t)) d^2N^0/dE dt
\ee

\nin where $P_d(t-t_d,t)$ is the probability that either a muon event or a
neutrino event ocurred in the interval from $t-t_d$ to $t$
and $d^2N^0/dE dt $ is the spectral rate without deadtime.
The probability, $P_d$,
 of an interaction  which causes dead-time is decomposed
as follows
$P_d = P_{d_\mu}P_{d_\nu}$,
using the poisson probability
that there were 0 events from $t-t_d \to t$, assuming a rate
$\Gamma$.

\beq
P_{d} =  1 - \exp{(-\Gamma{\delta t})}
\eeq

\nin For the muons $\Gamma_\mu = 3 Hz, \delta t = 35ms$.
We approximate the
neutrino induced dead time probability by
 evaluating the zeroth order contribution of
the neutrinos to $P_d$ at $t-t_d/2$, approximating $d^2N^0/dE dt $ as
constant during that short interval. Since $t_d \ll \tau_c$ this is a
good approximation.  Thus,
 $\Gamma_\nu \propto  dN_\nu^0$ and $\delta t = min(t,t_d)$.

The data in  our \ML code not specifically mentioned above, such
as fiducial detector volumes, the parameters of the detector resolution
functions, the energies and times of the background events in Kamiokande,
follow the treatment in \LL, and are not repeated here. We use the
standard likelihood function, see for example \cite{krauss:nucb, ll}.

The \MC program is a modified version of the one described in
\cite{krauss:nucb}. The parameters which describe the detector
efficiencies, resolution functions, the form of the neutrino
temperature etc, have been set to be identical to the ones for the
\ML analysis. This code (originally designed for O(1000) events) is
fairly sophisticated and includes the interactions of neutrinos other
than $\nueb$, and neutrino scattering from oxygen nuclei
 in the detector.  The additional types of
interactions in this code, and the more careful treatment of the
dominant reaction $\nueb p \to n e^+$, which includes nucleon recoil
effects, results in a small increase, $\sim 3\%$, in $<N>$ compared to
the
\ML code estimate. The difference is insignificant for the
SN1987A events however, as will
become obvious.

\nin {\bf Analysis of Results }

\noindent{\bf (a) Massive Neutrinos}

We first consider the limits on massive neutrinos.  The initial step
is to find the best fit \ML parameters for $\alpha, \tnuez , \tau_c ,
\tkii
$ and
$
\timb$ as a function of $\mnue$. The effect of this procedure is to
project the log likelihood onto the $\mnue$ axis. From this
projection, shown in Figure 1, we can find the 95\% confidence limit
from,

\be
\ln{ \cal L}^{Max} - .5 \chi_{dof}^2 (.05),
\label{eq:like}
\ee

\nin where we have 6 degrees of freedom ({\em dof}) for the
chi-squared distribution in the present instance. We will denote the
\ML value of a parameter by adding the subscript $L^+$, and the value
of a parameter at the 95\% confidence boundary with some or all (which
will be clear from the context) of the other parameters at their \ML
values by adding the subscript $L^-$. In this notation, from Figure 1,
\beq
(\mnue)_{L^+} &  = &  0 eV \\
(\mnue)_{L^-} &  = &  23 eV.
\eeq

\nin The likelihood function is extremely flat below $\mnue = 2eV$ so
this result does not strongly favor
an identically zero neutrino mass.
 Our value for $\unlikem$ is 8\% lower than the \LL
result $\unlikem = 25eV$, the entire difference being due to the
dead-time correction in our IMB likelihood function.

There is an additional constraint we may use in the analysis. Consider
Figure 2 wherein $\likekii$ and $\likeimb$ are shown as a function of
$\mnue$ (with ($\tau_c)_{L+}$ included for comparison). Note that
$\likekii$ reaches $4.2s$ at $\mnue= 23ev$ (where $\likeimb =
1s$).
  The \ML offset time for \Kam seems extraordinarily long,
especially in light of the fact that
 $\likekii$ exceeds $(\tau_c)_{L+}$ for $\mnue > 21.7 eV$.

To test our intuition in this regard, and to discover the acceptable
range of $t_{off}^{det}$ one can use our \MC code to
find, given $\likealpha, \liketnub$ and $ \liketau $, for a particular
$\mnue$, the probability, $\ptoffdet$, that the offset time in a
particular detector will not exceed a certain value.

   Note that we are interested in values of the mass parameter in the
range
$\likemnue (0 eV) < \mnue < \unlikemnue (23 eV)$.
We thus first determine, using the Maximum Likelihood method,
$\unlikekii$, the minimum acceptable $t^{kam}_{off}$,
 subject to the $\mnue$ constraint with all the other
parameters free. This is displayed in Figure 3. Then we use
our
\MC to construct $\ptoffkam$. If we find that $\unlikekii$ is ruled out
at the 95\% CL by this probability distribution,
 this then implies, since the likelihood function rules
out any smaller offset time, that the $\mnue$ corresponding to this
value is at least as unacceptable at this level.

We finally turn to the constuction of $\ptoffkam$.  Recall that
in our detector simulation we include dead-time for IMB and background
for Kamiokande.
Also note that the time of the first event depends (more strongly as the
number of events is decreased) on the number of events detected, and that
this number is not fixed by our \MC code, which temporally simulates the
neutrino burst and detection. Therefore we require \MC runs which result in
the desired number of events for each detector. Then we rank the
times of the first event of each such run and generate a cumulative
probability distribution for the time of the first event.
To improve the statistics, while conserving computer time, we choose a
range, $N = N^{det}\pm 1$, about the desired number of events.  (For this
purpose $N^{kam}=16$, including background, and $N^{imb}=8$.)
( Both detectors,
for the parameters of interest here, have a flat distribution for
the expected time of the first event in the neighborhood of the
number of events actually observed. )
 We use 1000 \MC runs of each
detector to acquire the data for the construction of $\ptoffdet$.
 Typically
about $20 - 30\%$ of the
\MC runs fall in the accepted range of $N^{det}$.

In Figure 4a we plot $P(t_{off}^{det} < t
|N^{det}, 21eV)$.  Also shown are $\likekii$, and $\likeimb$, for
$\mnue=21eV$.   Note that while $\likeimb =.9s$ is
located near the mean of the distribution, $\likekii =3.8s$ is
in the tail. In Figure 4b $P(t_{off}^{Kam} < t
|N^{Kam}, 19.6eV)$ is plotted with $\likekii$ and $\unlikekii$
indicated.
Since
$\unlikekii > t^* $, where $t^*$ is defined by
$P(t_{off}^{kam} < t^*|N^{kam}, 19.6eV) = .95$,
 $\mnue \ge 19.6eV$ is excluded at the
95$\%$ confidence level.

We thus find an additional decrease of $15 \%$ in the $\mnue$ upper
limit derivable using the \MC generated probability $\ptoffkam$ in
addition to the \ML procedure.  This is a significant factor, and
further underscores the utility of Monte Carlo simulation of the
data.  When combined with the effect of incorporating deadtime
in the IMB
detector, which fortuitously plays a significant role because of the
paucity of observed events in Kamiokande, we have been able to reduce the
upper limit on the $\mnue$ mass by over 25$\%$ compared to previous
analyses.
\vskip 0.1in

\noindent {\bf (b) Vacuum Mixing and Nearly Massless Neutrinos}

Next we turn to our results for vacuum mixing. In this case the
\ML values for the offset times are 0 for all
values of $\svac$, therefore we do not have to \MC the neutrino
burst. In Figure 5 $\ln {\cal L} (\svac)$ is displayed for the range
$0 \le \svac \le 1$. (We ran the \ML code with and without including
the offset times. The difference in $\ln {\cal L}$ never exceeded
.02\%.  Thus the offset times are irrelevant parameters, and we consider
only 5 {\em dof}.) In Figure 5 the likelihood
function peaks at $\svac=.45$.  However, the likelihood function is
relatively flat over the entire range so non-zero mixing is
only marginally preferred.  The likelihood ratio
 for $\svac=0.45$ compared to $\svac=0$
is 5.5.

Also of interest, perhaps to supernovae model builders, is our \ML
extraction of neutrinosphere temperatures.  In Figure 6a-c we display
$\tnue, \tnueb$ and $\tnum$ for $\svac$ = 0, .45 and 1 respectively.
The main feature is that in all cases $\tnueb \to \tnum$ gradually,
with $\tau_2$ = 9.13, 8.72 and 8.35 respectively. This long timescale is
something of a surprise.

A final question is whether the
admission of constraints on
\SN parameters, other than those purely obtainable from the
neutrino data alone, would allow one to further limit $\svac$.  In
Figure 7a-c we show: the \ML neutron star binding energy, $\Ebind$, in
units of $10^{53}$ ergs; the intitial electron neutrino temperature,
$\tnuez$, in MeV; and the cooling time-scale, $\tau_c$, in
seconds. The entire range of the latter seems acceptable based on
estimates from supernova models. One may ask whether constraints such as
$\Ebind < 4.5 $ or $\tnuez > 3 MeV$, would limit $\svac$.
 From Figure 7 it appears that this could be the case.

To address this question we find the 95\% confidence level regions in the
$\Ebind$ , $\tnuez$ plane for several values of $\svac$. The remaining
parameters are all permitted to find their \ML values\footnote{In our
formalism $\Ebind$ is determined by the combination of $\alpha$, $\tnuez$
and $\tau_c$, thus in the present context we obtain $\alpha$ from the fixed
values of $\Ebind$ and $\tnuez$ and the free value for $\tau_c.$}.  In
Figure 8 we display the results for 6 values of $\svac$ from .1 to 1. The
outer contour in each box is the 95\% confidence limit (5 degrees of
freedom), while the inner contours; 50\%, 25\%, and 10\% C.L.\ ,
are shown to allow the reader to assess the
 character of the surface. It is apparent from Figure
8 that no reasonable $\Ebind$ provides a further solid constraint on
$\svac$.

We have constructed Table 1 to present the maximum allowed $\svac$ at
95\% confidence level in terms of a given minimum permissible
$\tnuez$, which we designate $\tnuezm$; this is understood to refer to
a limitation on the electron neutrino temperature provided
independently of the neutrino data, such as may arise from supernovae
modeling.  If one cannnot bound $\tnuez$ from below then the parameter
$\svac$ cannot be constrained. In order to rule out large vacuum angle
solutions to the solar neutrino problem ($.7 \le \svac \le .9$
\cite{justso}), using the \SN neutrino data, a rigorous argument that
supernovae dynamics {\em require} $\tnuez >$ \hbox{ 4 MeV} appears to
be needed.  The $\tnuez$ parameters in Table 1 are well within the
typical range of 3-5 MeV in the supernovae model literature
\cite{tnubar,tnue}.

\vskip .1in
\nin {\bf Conclusions}
\vskip .1in

Our results demonstrate several important lessons for statistical
analyses of constraints on neutrino properties from a nearby supernova
neutrino burst, as well as refining these constraints for the observed
burst from SN1987A.  In the first place, while a \ML procedure can
provide very powerful constraints on model parameters, it alone cannot
address the question of whether these model parameters can be
realistically achieved.  When these parameters have to do with
features of the observed burst, and not internal features of an
underlying supernova model, then a \MC procedure such as we have
devised \cite{krauss:nucb} can prove to be very useful in further
strengthening constraints on neutrino properties.

Next, we have seen that the ability of the SN1987A burst signals in
\Kam and IMB to constrain a non-zero electron neutrino mass is in some
sense fortuitous, due to the ``tension'' of the \Kam and IMB
data---in particular the apparent paucity of events in \Kam relative to
IMB.  For this reason, when we included deadtime in IMB we were able to
further extend the lever-arm in constraining $\mnue$.  Our final
result, $\mnue < 19.6 eV$ is approximately 25\% stronger than
the previous best limit.

Finally, we find that the ability of the combined SN1987A neutrino
bursts to constrain neutrino masses does not at present extend to an
ability to constrain neutrino mixing angles in any model independent
way. In particular, because of the uncertainty in the timescale for
neutronization, without introducing strong model
dependence---in particular contraints on $\Ebind$ and $\tnueb$---one
cannot limit $\svac$ from the \SN neutrino events.  This argues
against the claim made in \cite{smirnov}. It will be interesting to
determine just how strong the constraints might become for a galactic
supernova burst, and this issue is currently under investigation.

\vskip .1in
We thank Steve Dye, Robert Svoboda and Martin White for useful
conversations. We also thank the IMB collaboration for providing
unpublished data.

\newpage

\nin {\Large \bf Figure Captions}
\bigskip

\nin Figure 1:
The projection of the log Maximum Likelihood onto the $\mnue$ axis.
\bigskip

\nin Figure 2:
The \ML offset times for the IMB and Kamiokande detectors as $\mnue$
is varied. For comparison the \ML supernovae cooling timescale is
also shown.
\bigskip

\nin Figure 3:
The log likelihood as a function of the offset time in the Kamiokande
detector for several values of $\mnue$. The horizontal line indicates
the 95\% C.L. boundary.
\bigskip

\nin Figure 4:
Shown are Monte Carlo generated cumulative probability
distributions for the time of the first event in a detector given the
supernovae model parameters. In (a) the neutrino mass is 21 eV,
and the distributions for IMB and Kamiokande are shown.  The \ML offset
times are also indicated. In (b) the neutrino mass is 19.6 eV. The
distribution for Kamiokande is shown, as are the \ML and 95\% C.L. offset
times. The short horizontal line is at P=95\%.
\bigskip

\nin Figure 5:
The projection of the log Maximum Likelihood onto the $\svac$ axis.
\bigskip

\nin Figure 6:
Temporal profiles of
the \ML electron, anti-electron, and muon neutrinosphere temperatures
for mixing angles of $\svac=0$ (a),
$\svac=.45$ (b) and $\svac=1$ (c).
\newpage

\nin Figure 7:
As a function of neutrino mixing angle, the \ML
 neutron star binding energy in units of $10^{53}$ ergs (a),
 initial electron neutrinosphere temperature in MeV (b), and
 supernovae neutrinosphere cooling timescale in seconds (c).
\bigskip

\nin Figure 8:
\ML Projections into the plane of the neutron star binding energy and
initial electron neutrinosphere temperature for several choices of
the neutrino mixing angle. The binding energy is in units of $10^{53}$
ergs and the temperature in MeV. The contours are displayed at
the 95\%, 50\%, 25\% and 10\% Confidence Levels. For $\svac=1$ the
10\% C.L. contour does not exist. For $\svac=.1$ the
25\% and 10\% C.L. contours do not exist. (See Figure 5).

\vskip 0.3in

\nin \bc
{\bf Table 1 : Maximum $\tnuez$ on the 95\% C.L. boundaries of Figure 8} \\
\vskip 0.1in
\begin{tabular}{||c|c||} \hline
$\tnuezm $ {\small (MeV) }& $\svac^{95}$ \\ \hline \hline
3.59 & 1.0 \\
3.67 & 0.9 \\
3.96 & 0.7 \\
4.27 & 0.5 \\
5.00 & 0.25\\ \hline
\end{tabular}
\ec


\newpage

\begin{thebibliography}{99}

\bibitem{imb}
C.B.Bratton \etal , Phys. Rev. D {\bf 37}, 3361 (1988)
\\ R.M.Bionta \etal , Phys. Rev. Lett. {\bf 58}, 1494 (1987)

\bibitem{kam}
K.S.Hirata \etal , Phys. Rev. D {\bf 38}, 448 (1988) \\
K.S.Hirata \etal , Phys. Rev. Lett. {\bf 58}, 1490 (1987)

\bibitem{krauss:nature}
L.M.Krauss, Nature {\bf 329}, 689 (1987)

\bibitem{krauss:nucb} L.M.Krauss, P.Romanelli, D.Schramm, and
R.Lehrer, Nucl. Phys. B. {\bf 380}, 507 (1992)

\bibitem{ll}
T.J.Loredo, and
D.Q.Lamb, Ann. N.Y. Acad. Sci. {\bf 571}, 601 (1989)

\bibitem{sd}
Steve Dye, private communication

\bibitem{walker}
L.F.Abbot, A.De R\'{u}jula, and T.P.Walker, Nucl. Phys. B. {\bf 299},
734 (1988)

\bibitem{smirnov}
A.Y.Smirnov, D.N.Spergel, and J.N.Bahcall, preprint hep-ph 9305204 (1993)

\bibitem{justso}
P.I.Krastev and S.T.Petcov, Phys. Rev. Lett. {\bf 72}, 1960 1994 \\
P.J.Kernan, Ph.D.Thesis Ohio State University (1993)
\\ S.L.Glashow
and L.M.Krauss, Phys.Lett. {\bf B190} 199 (1987) \\ V.Barger,
K.Whisnant and R.J.N.Phillips, Phys. Rev. D { \bf 24} 538 (1981) \\
J.N.Bahcall and S.C.Frautschi, Phys. Lett. {\bf B29}, 263 (1969) \\
B.Pontecorvo, Sov.JETP, {\bf 26} 984 (1968)


\bibitem{snp}
An excellent
review is found in J.N.Bahcall, {\em Neutrino Astrophysics} (Cambridge
University Press, Cambridge, 1989).

\bibitem{pdg}
Particle Data Group, {\em Phys.Rev.}{\bf
D50} 1173 (1994)

\bibitem{kolb}
E.W.Kolb, A.J.Stebbins, and M.S.Turner, Phys. Rev. D.
{\bf 35}, 3598 (1987)

\bibitem{tnubar}
R.Mayle, J.R.Wilson, and D.N.Schramm, Ap.J. {\bf 318}, 288 (1987)


\bibitem{tnue}
A.Burrows, Ann.Rev.Nucl.Sci. {\bf 40}, 181 (1990)

\end{thebibliography}
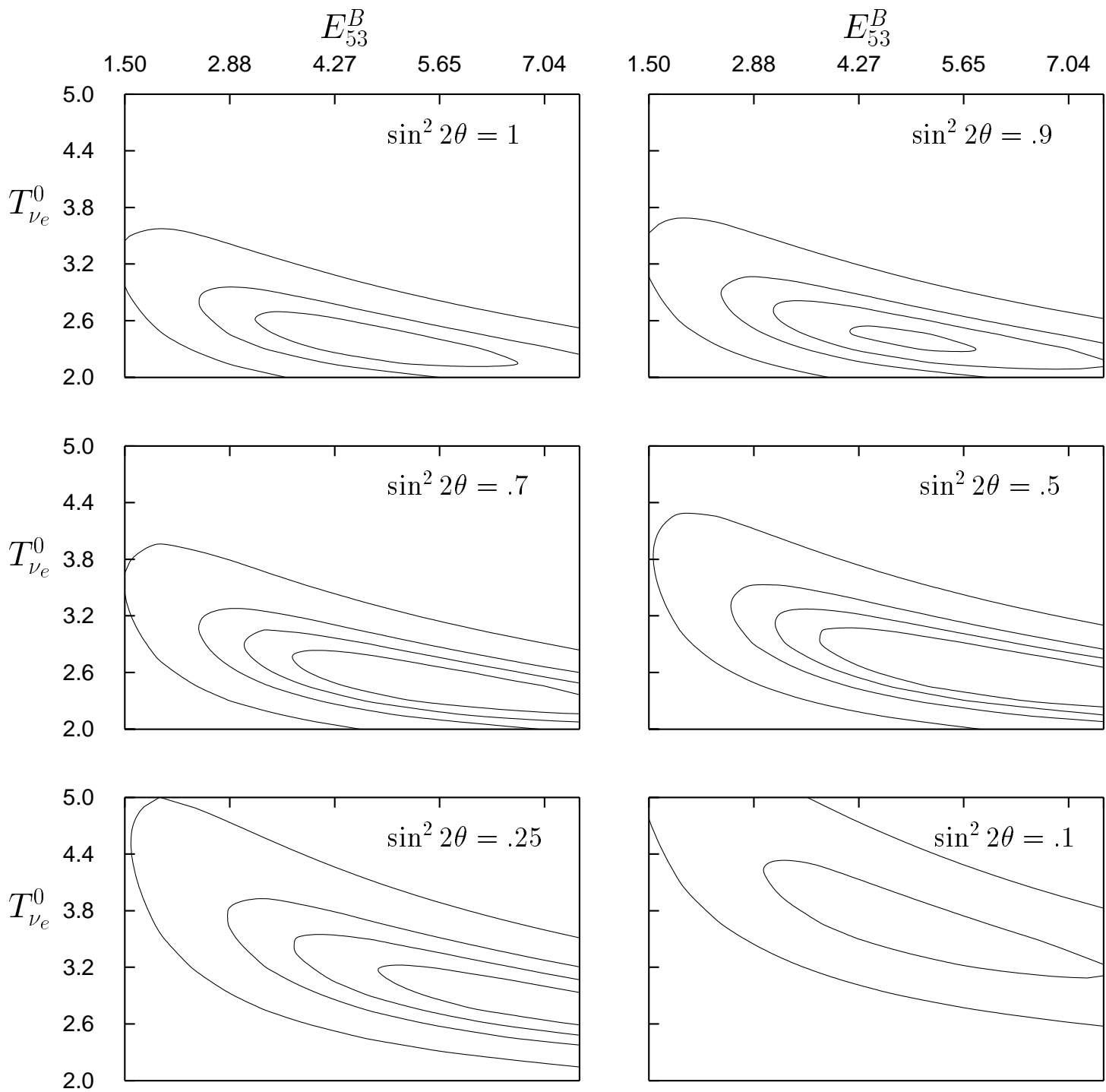
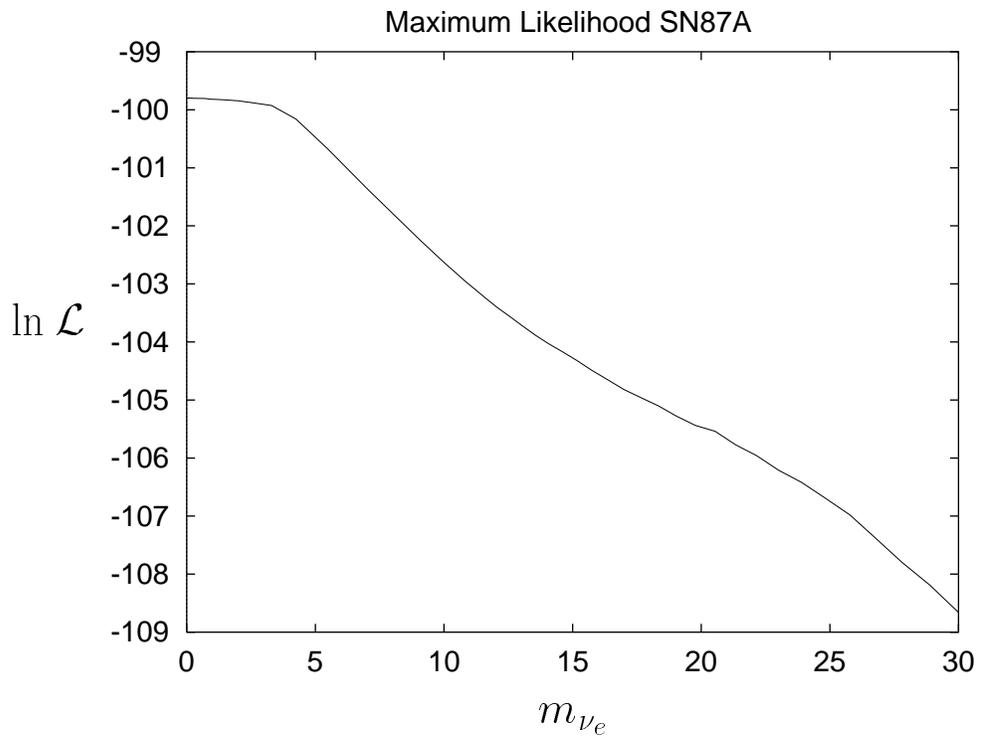
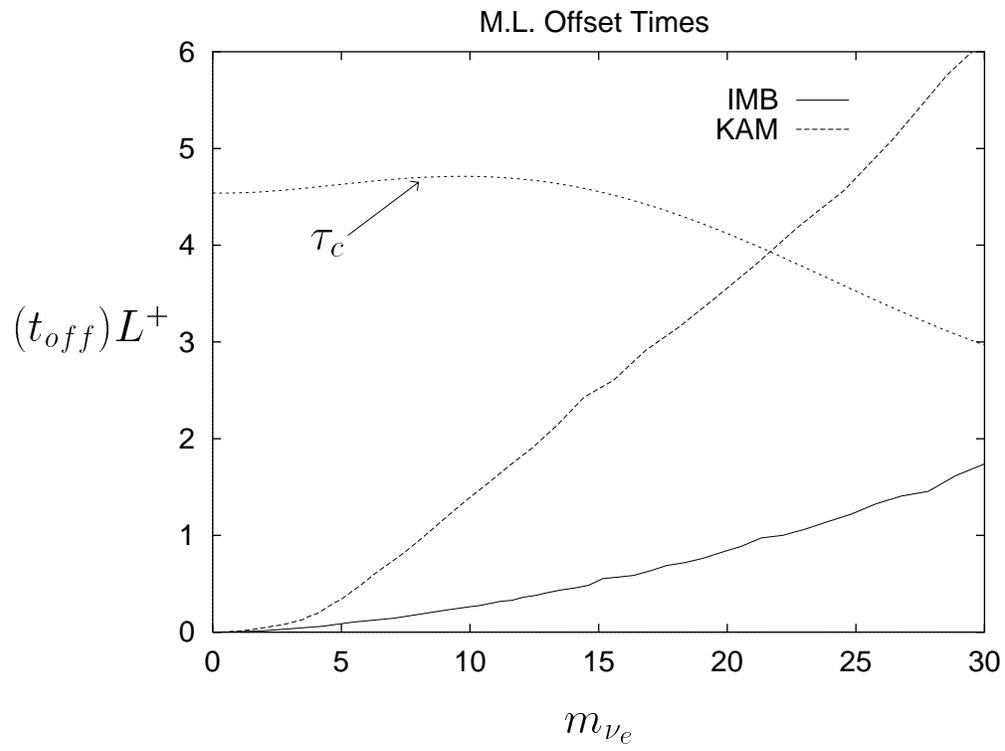
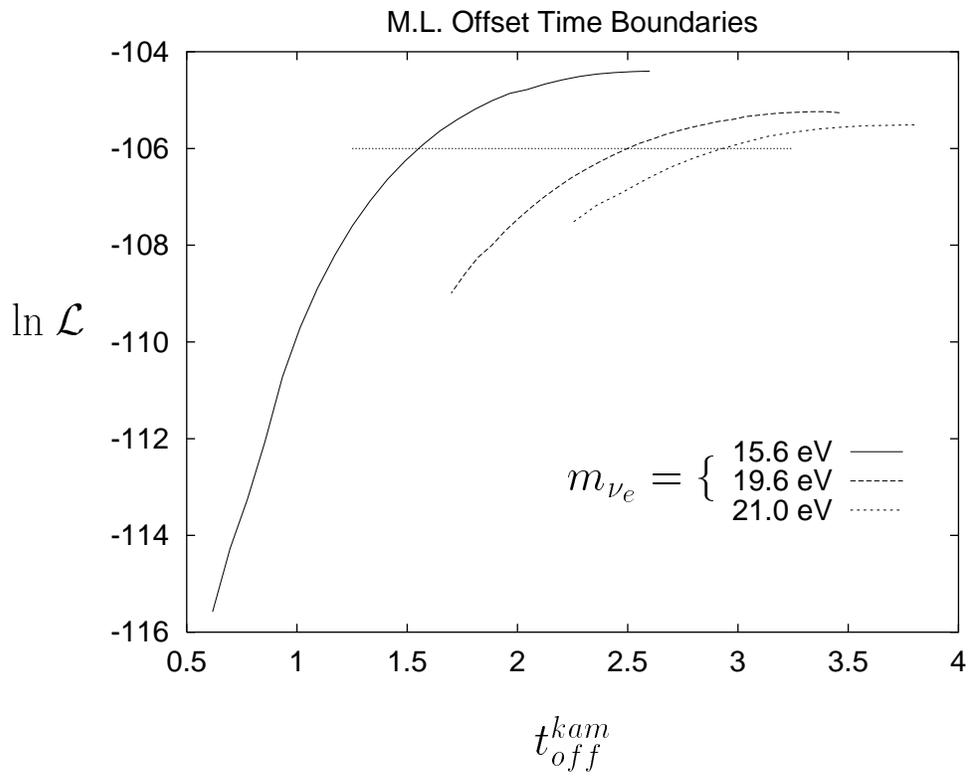
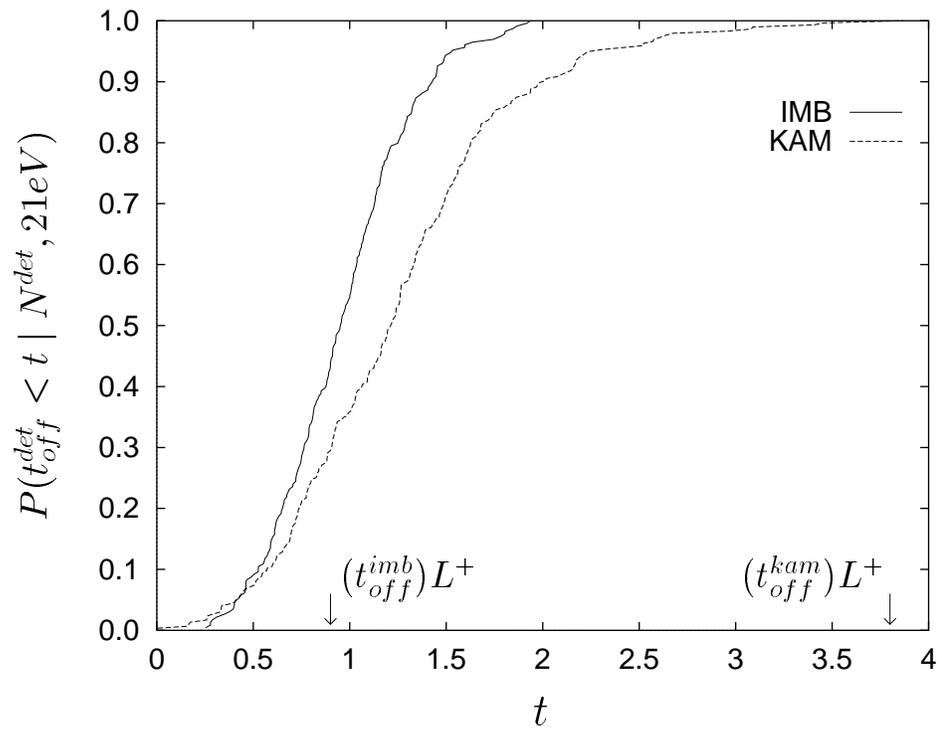
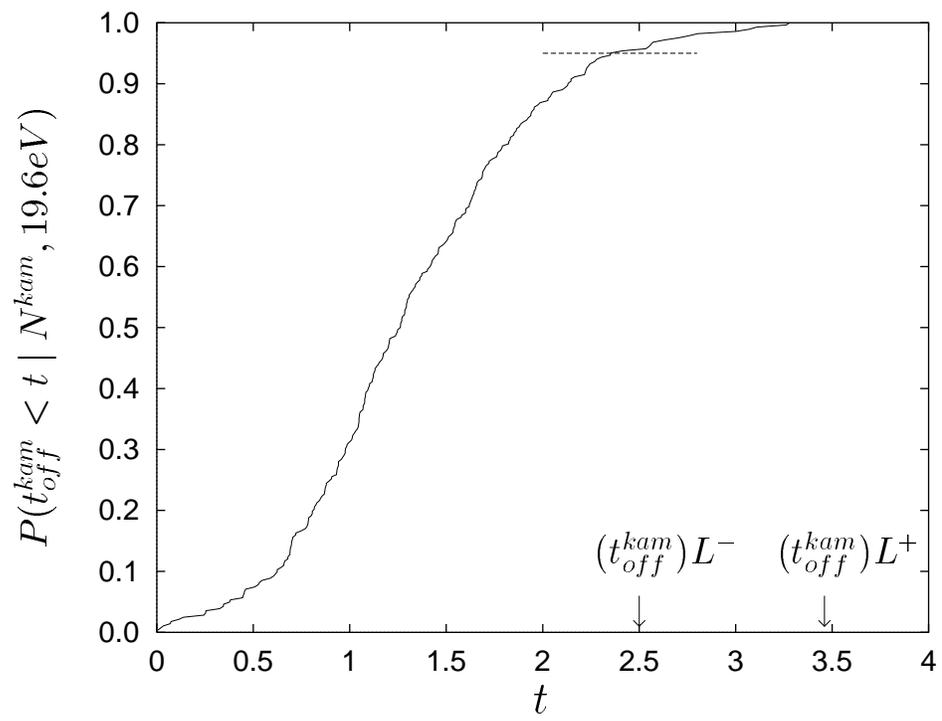
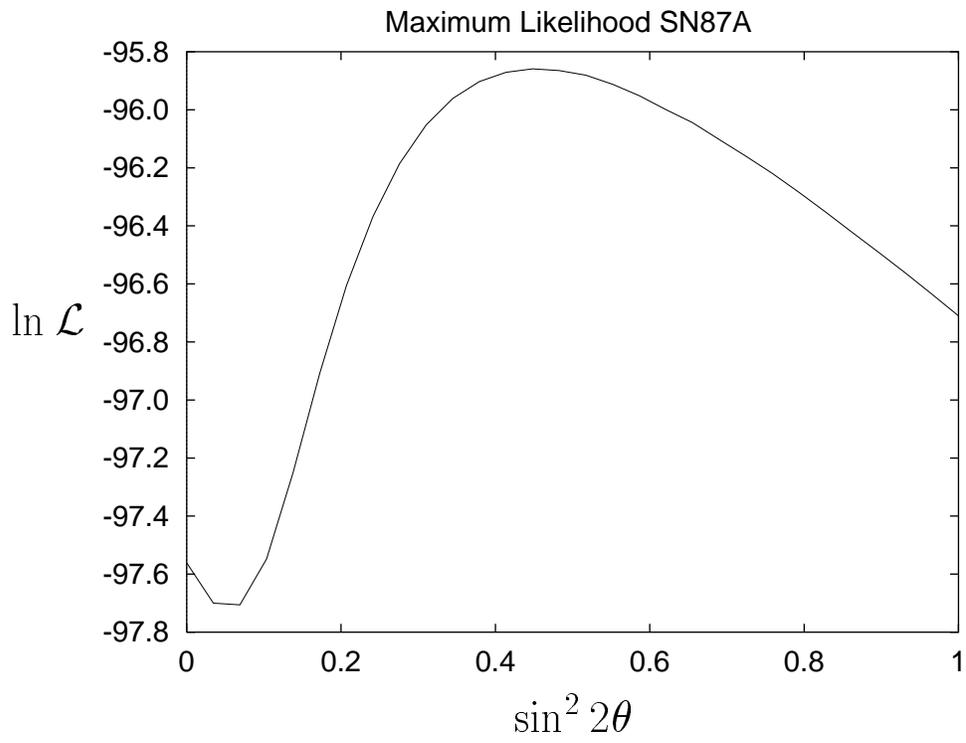
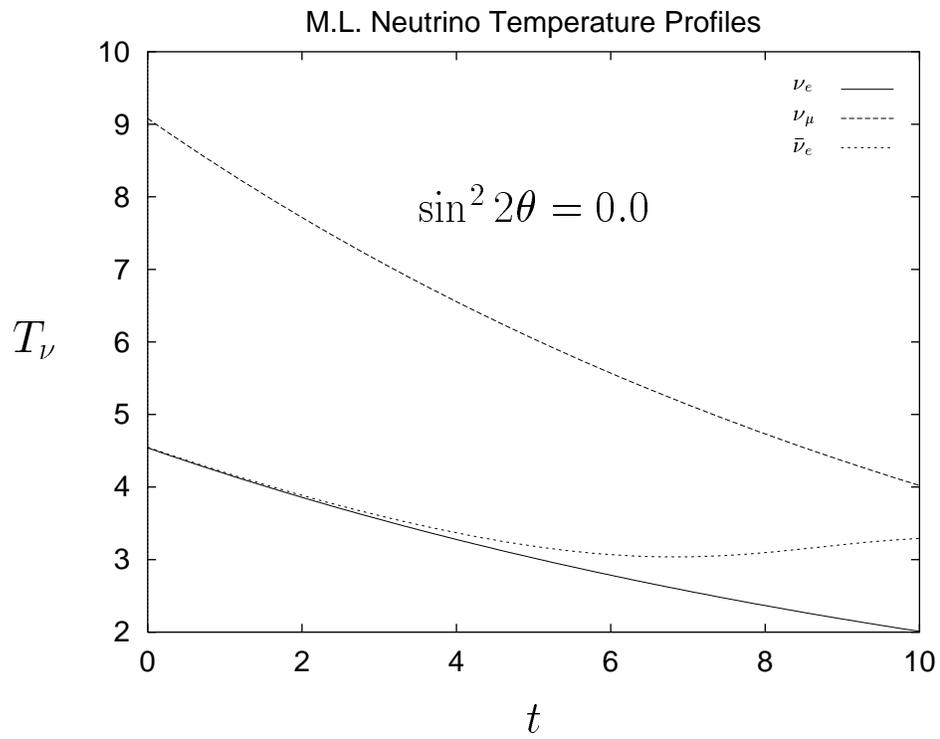
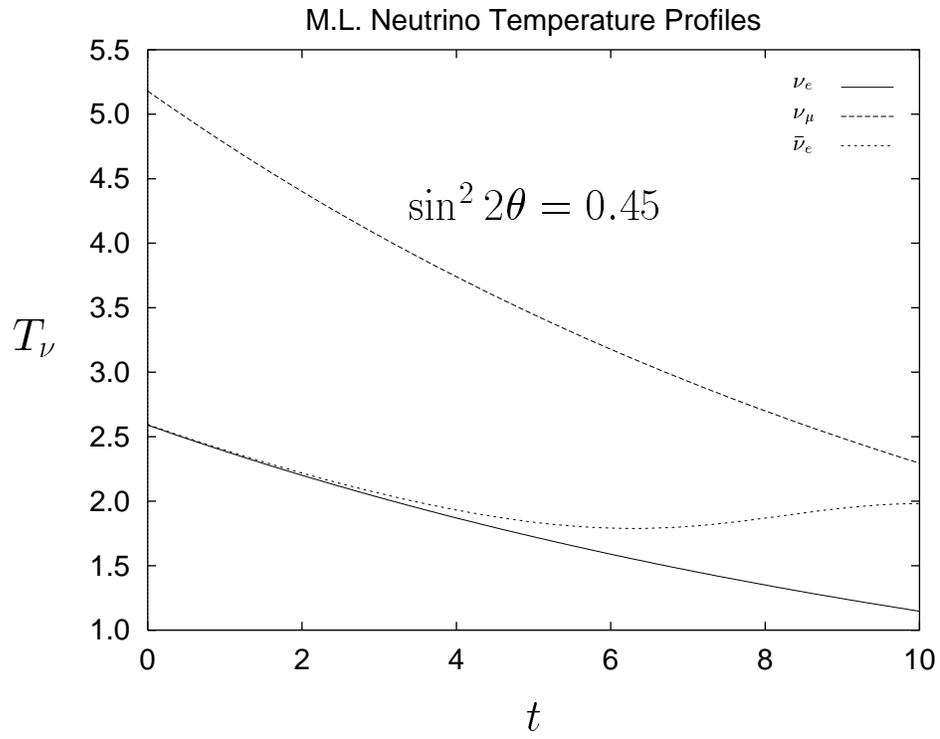
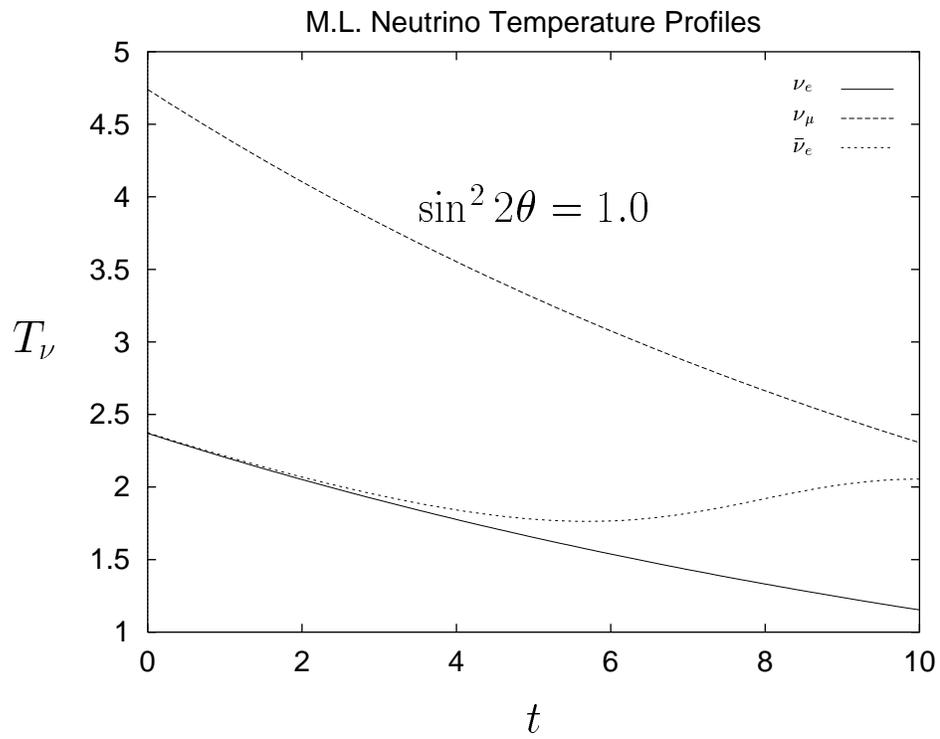
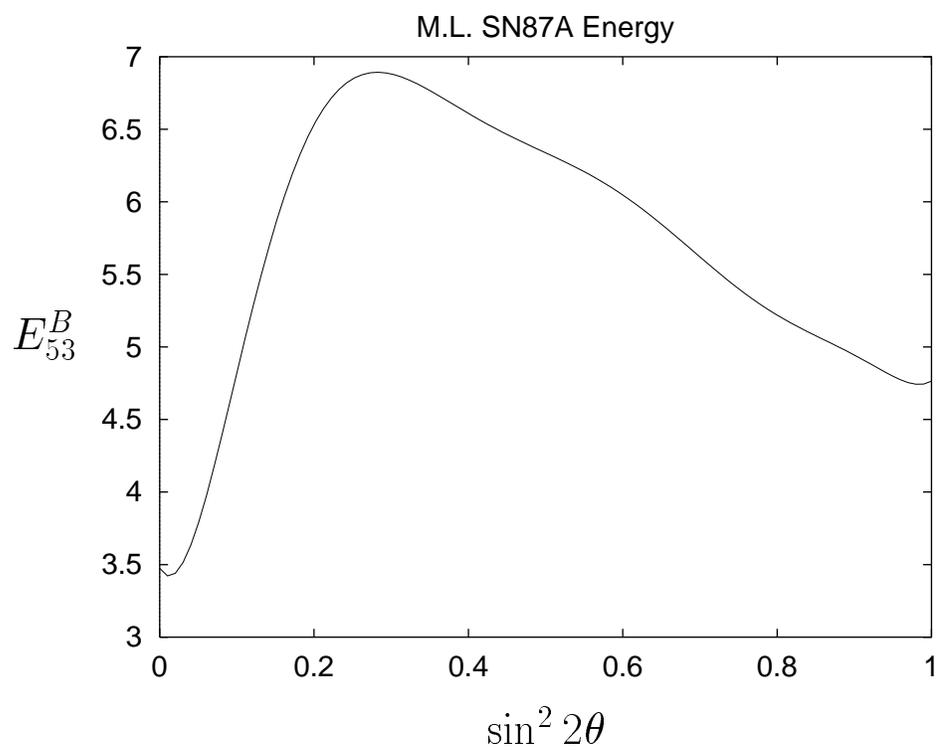
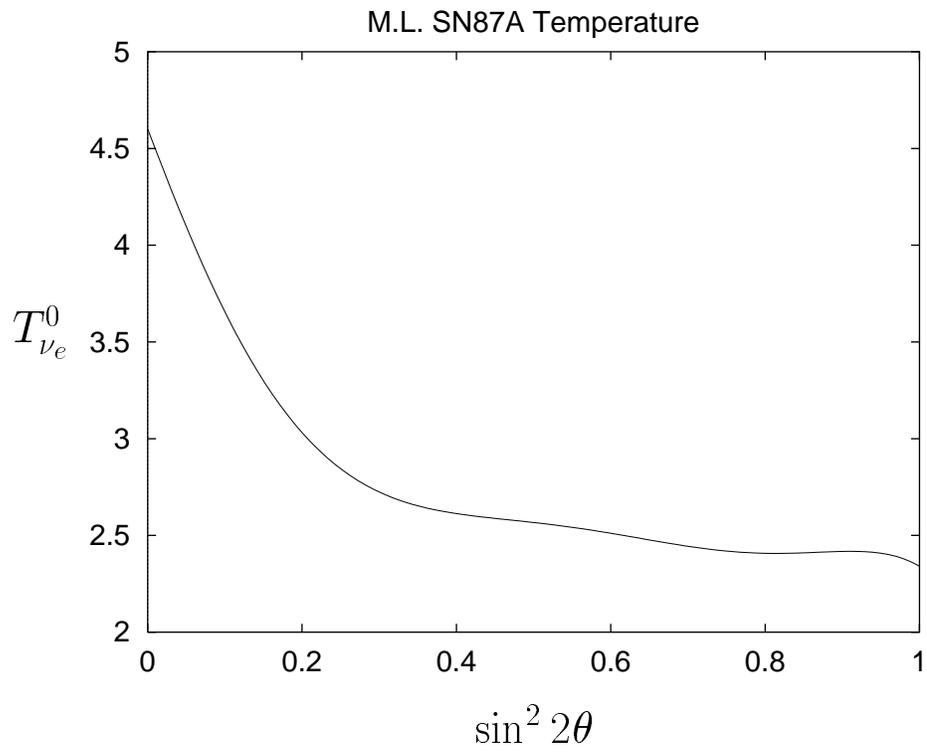
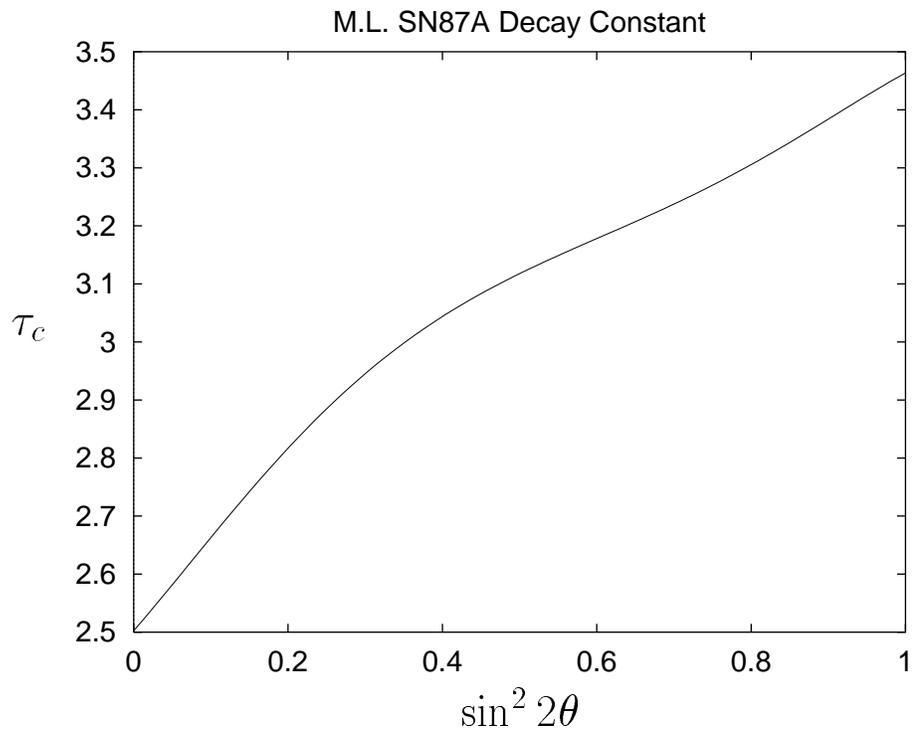
\end{document}